\documentclass[aps,prl,reprint,showpacs,floatfix]{revtex4-1} %

\usepackage{amsmath,amssymb,amsfonts}
\usepackage{graphicx}

\begin{document}

\title{Understanding deviations from ray optics at curved interfaces}

\date{\today}

\author{Pia Adam}\email[Corresponding author: ]{pia.adam@tu-ilmenau.de}
\author{Jakob Kreismann}
\author{Martina Hentschel}
\affiliation{Institute for Physics, Technische Universit\"at Ilmenau, Weimarer Stra\ss{}e 25, 98693 Ilmenau, Germany}

\begin{abstract}
Ray optics is a useful tool even in the regime where, actually, full wave-calculations would be appropriate. However, wave-inspired adjustments are needed to ensure the accuracy of ray-based predictions.
These corrections are known as the Goos-H\"anchen shift, a lateral shift along the interface, and the Fresnel filtering effect, an angular shift, that violates Snell's law and the principle of ray-path reversibility. Whereas they are well established at planar interfaces, an accurate description of microlasers and other microoptics devices requires their precise knowledge at the curved boundaries characteristic for these devices. 
Here, we present analytical and numerical results that highlight the role of boundary curvature and show the clear deviations from the planar case.
We introduce an intuitive picture that allows for a straightforward understanding why Fresnel filtering grows considerably with curvature whereas the Goos-H\"anchen shift becomes less important. 
\end{abstract}


\maketitle

Ray optics can both explain natural phenomena as the rainbow and is an easy and versatile tool to describe microscale optical systems like microlasers \cite{Vahala,McCall1992} and their farfield emission \cite{Yan2009,Lee2007,limacon,Schwefel2004}.  
However, just as geometric optics cannot explain the wave-originating supernumerary rainbow, corrections to ray optics are needed when reducing the cavity size to the order of several light wavelengths. 

\begin{figure}[b!]
 \includegraphics[width=.47\textwidth]{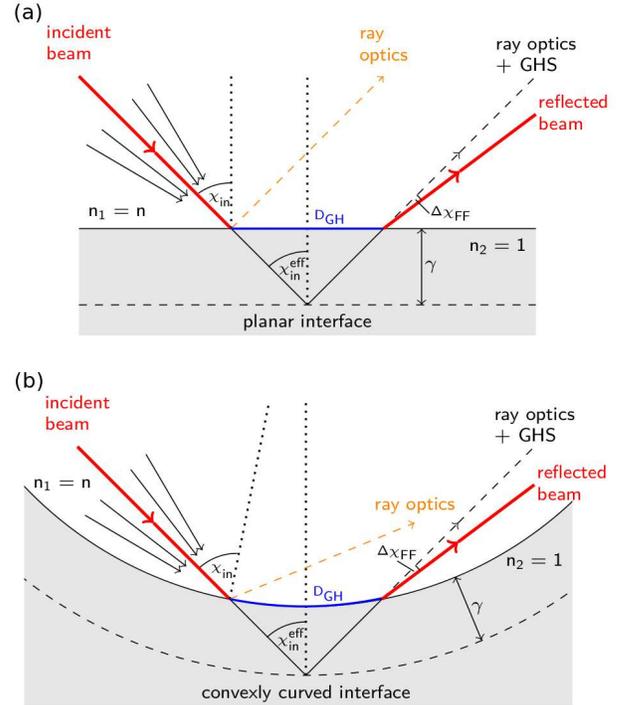}
 \caption{\label{fig:Prinzipskizze}(color online). Schematic picture of the beam shifts at (a) planar, (b) curved interfaces. Incident and reflected beam are shown in red. The relative refractive index is $n=n_1/n_2$. The effective interface is depicted as dashed line, $\gamma$ is the penetration depth. The normals to the interface and the effective interface (dotted lines) define the angle of incidence $\chi_\text{in}$ and the effective angle of incidence $\chi_\text{in}^\text{eff}$. The reflected beam predicted from ray optics is shown as dashed orange arrow. The Goos-H\"anchen shift $D_\text{GH}$ (blue) is determined by ray optics at the effective interface. The reflected beam deviates by an angle $\Delta\chi_\text{FF}$ from the GHS-corrected beam (dashed arrow), this angular deflection is the Fresnel filtering.}
\end{figure}

Corrections to ray optics have been long known, namely in the form of beam shifts: In the 1940s, Fritz Goos and Hilda H\"anchen observed deviations in the total reflection from a series of glass plates \cite{GoosHaenchen}.  This Goos-H\"anchen shift (GHS) at a planar interface, depicted in Fig.~\ref{fig:Prinzipskizze}(a), occurs for total internal reflection (angle of incidence $\chi_\text{in}$  larger than the critical angle $\chi_c = \arcsin 1/n$ where $n=n_1/n_2$ is the relative refractive index). It results from an interference effect  \cite{Artmann} and yields a lateral shift of the order of the vacuum light wavelength $\lambda$ between incident and reflected beam \cite{HorowitzTamir,Lai1986,Lotsch1970}. For an intuitive understanding, recall the finite penetration depth $\gamma$, of the order of $\lambda$, of totally reflected light into the optically thinner medium. The reflection can thus be assumed to occur at an effective interface (dashed line in Fig.~\ref{fig:Prinzipskizze}) under an effective angle of incidence $\chi_{\text{in}}^{\text{eff}}$ \cite{SchomerusHentschel_Fresnel_laws_curved,KotikHentschel}.
Whereas  $\chi_\text{in}^\text{eff} = \chi_\text{in}$ holds at a planar interface, we find $\chi_{\text{in}}^{\text{eff}}<\chi_{\text{in}}$ in the convex case, cf.~Fig.~\ref{fig:Prinzipskizze}(b), which is the typical case for light in a microcavity.

Before we discuss the consequences, we introduce the second beam shift effect, the Fresnel filtering (FF) \cite{TureciStone02} (or angular Goos-H\"anchen shift, \cite{Merano2009,Fiedler-Ferrari1991,Aiello_overview,SchomerusHentschel_phase-space,GoetteShinoharaHentschel2013}).
Its origin is the finite spatial extent of a light \emph{beam} (a solution of Maxwell's equations) as opposed to the light \emph{rays} used in geometric optics. 
This finite beam width induces a distribution of angles of incidence around a mean $\chi_\text{in}$. This distribution will, moreover, strongly depend on the interface curvature.
Now suppose $\chi_\text{in} = \chi_c$. Then, all smaller angles in the distrubtion will be partially transmitted, whereas all larger angles will be totally reflected -- thus the mean angle of the reflected light will deviate from $\chi_\text{in}$: We are confronted with a violation of Snell's law,
$\chi_\text{out} = \chi_\text{in} + \Delta\chi_\text{FF}$, with the FF correction $\Delta\chi_\text{FF}$. FF will be most important around $\chi_c$ and for beams narrow in space, i.e., with a broad angular distribution. 

Here, we will show that both GH and FF are sensitive to boundary curvature, and do so in opposite ways. The paper is organized as follows. We first present numerical results of the reflection of a light beam at a curved interface and confirm them by analytical calculations. Then, we investigate the curvature dependence of GH and FF and end with an explanation of our results in an easy-to-use picture and discuss the impact of our findings.

\begin{figure}
 \includegraphics[width=.47\textwidth]{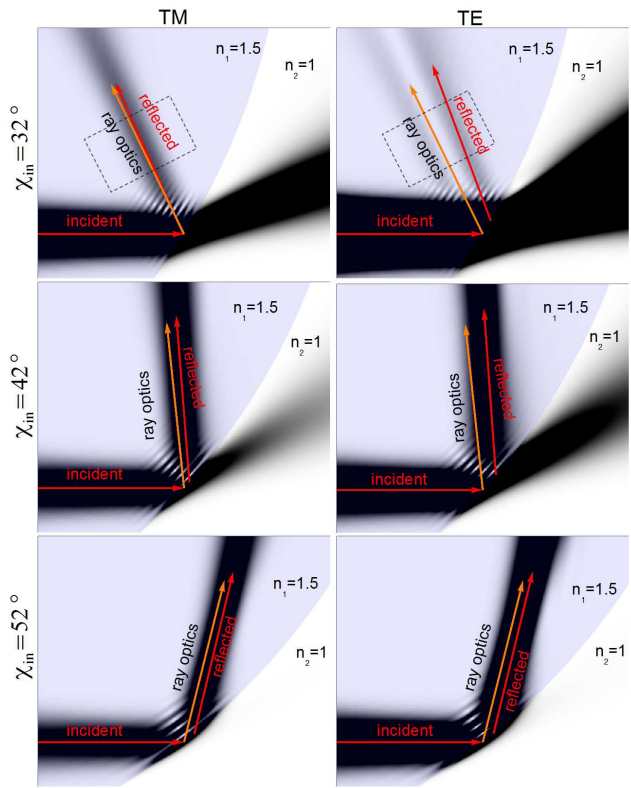} 
 \caption{\label{fig:panel_numerics}(color online). Real space pictures of a Gaussian beam reflected at a convex interface. The panels show TM and TE polarization (left and right column, respectively) and angles of incidence near the Brewster angle, near the critical angle, above the critical angle (top to bottom). The direction of incident and reflected beams are given by red arrows, the direction of the reflected beam expected from ray optics is given by the orange arrow. The dashed rectangle in the first two pictures depicts the window used to determine the direction of the reflected beam. For all panels $kR=400$, $n=1.5$, $w/\lambda=5$}  
\end{figure}

Numerical results are obtained from full electromagnetic calculations simulating a Gaussian beam incident on a curved (convex or concave) dielectric interface. These simulations have been performed with the finite-difference time-domain (FDTD) method, using a freely available software package \cite{meep}. Figure \ref{fig:panel_numerics} shows the results of such FDTD simulations for the reflection of a Gaussian beam at a convexly curved interface for the two polarizations, transverse magnetic (TM, left) and transverse electric (TE, right), for subcritical, near-critical, and supercritical incidence (from top to bottom).  The ray propagation directions of the incident and reflected beams were obtained as mean value of the Poynting vectors taken in a suitably chosen window across the beams (dashed rectangle in the upper panels of Fig.~\ref{fig:panel_numerics}). The resulting propagation directions are denoted by red arrows in Fig.~\ref{fig:panel_numerics}, whereas the geometric optics results is marked by the orange arrow. The deviation is clearly visible.  
The beams shifts, $D_{\text GH}$ and $\Delta \chi_{\text FF}$, are calculated as difference between the simulated reflected beam and its ray-optics expectation. 

Above $\chi_c$, GHS dominates in both polarizations. Near the critical angle, the FF effect is clearly visible and is, as expected, most important here.
We anticipate the least deviations between ray and wave description for subcritical incidence, and this holds indeed for TM polarization. The TE result comes as surprise on first sight, but is readily understood due to the Brewster angle, $\chi_B = \arctan 1/n < \chi_c$. No TE polarized light is reflected at $\chi_B$, corresponding to the gap in the reflected beam.

Analytical calculations for GHS and FF are performed for convex (and planar) interface boundaries using the appropriate Fresnel reflection coefficients.  The beam shifts are obtained as expectation values of the electromagnetic field \cite{Porras_moment-method}.  
As most devices have a locally constant curvature, $\kappa = 1/R$, it is sufficient to calculate the corrections for fixed radius of curvature $R$. 
Consequently, we work in polar coordinates $(r,\alpha)$, with the disk center as origin. We express the incident light beam at the interface, $r=R$, in cylinder functions with angular wavenumber $m$ (Fourier-Bessel decomposition), $E_I(\alpha) = \sum_m e_I(m) e^{im\alpha} J_m(nkR)$ with the vacuum wavenumber $k=2\pi/\lambda$, the Bessel function of the first kind $J_m(z)$ and the beam profile in angular momentum space $e_I(m)$.

To compare the results from this analytical approach with the FDTD results, the transverse beam profile in angular momentum space $e_I(m)$ is chosen to be a Gaussian centered at $m_0$ with standard deviation $\sigma$,
\begin{equation}
 e_I(m) = \frac{1}{\sqrt{2\pi}\sigma} e^{-\frac{(m-m_0)^2}{2\sigma^2}}\;.
 \label{eq:gaussian}
\end{equation}
In the numerical simulations, the corresponding real space profile of width $w/\lambda = \sqrt{2} k R / (\pi \sigma)$ is implemented. 

Angular momentum conservation in this radially symmetric geometry yields a relation between the angular wavenumber $m$ of the cylinder function $J_m$ and the angle of incidence $\chi$ \cite{SchomerusHentschel_Fresnel_laws_curved}, 
\begin{equation}
 \sin(\chi) = \frac{m}{nkR}.
 \label{eq:angle_of_incidence}
\end{equation}
Modeling a ray with angle of incidence $\chi_0$, thus, means that we have to chose an incident beam profile $e_I(m)$ which is strongly peaked at the central angular wavenumber $m_0$ corresponding to $\chi_0$ like the Gaussian profile of Eq.~\eqref{eq:gaussian}.

The reflected light beam is then given by $E_R(\alpha) = \sum_m e_R(m) e^{im\alpha} J_m(nkR)$ where $e_R(m)=\rho(m)e_I(m)$ is the reflected beam profile that is weigthed by the Fresnel reflection coefficient $\rho(m)$ for curved (rather than planar) interfaces provided in \cite{SchomerusHentschel_Fresnel_laws_curved}. 

The beam shifts can be defined as expectation values of the reflected beam \cite{Porras_moment-method}. 
The GHS is easiest obtained from the expectation value of the polar angle
\begin{equation}
 \Delta\alpha_\text{GH} = \langle \alpha \rangle = \frac{\int_{-\pi}^{\pi}\! \text{d}\alpha \, E_R^*(\alpha) \alpha E_R(\alpha)}{\int_{-\pi}^{\pi}\! \text{d}\alpha \, E_R^*(\alpha) E_R(\alpha)}.
 \label{eq:definition_GH}
\end{equation}
The lateral shift $D_{\text GH}$ along the interface, given in multiples of the wavelength $\lambda$, is then $D_\text{GH}=nkR\,\Delta\alpha_\text{GH}/2\pi$.

The FF correction $\Delta\chi_\text{FF}$ can be directly translated into a shift in the central angular wavenumber component $m_0$ via Eq.~\eqref{eq:angle_of_incidence},
\begin{equation}
 \Delta m_\text{FF} = \langle m-m_0 \rangle = \frac{\sum_m e_R^*(m)(m-m_0)e_R(m)}{\sum_m e_R^*(m)e_R(m)}.
 \label{eq:definition_FF}
\end{equation}

\begin{figure} 
  \includegraphics[width=.47\textwidth]{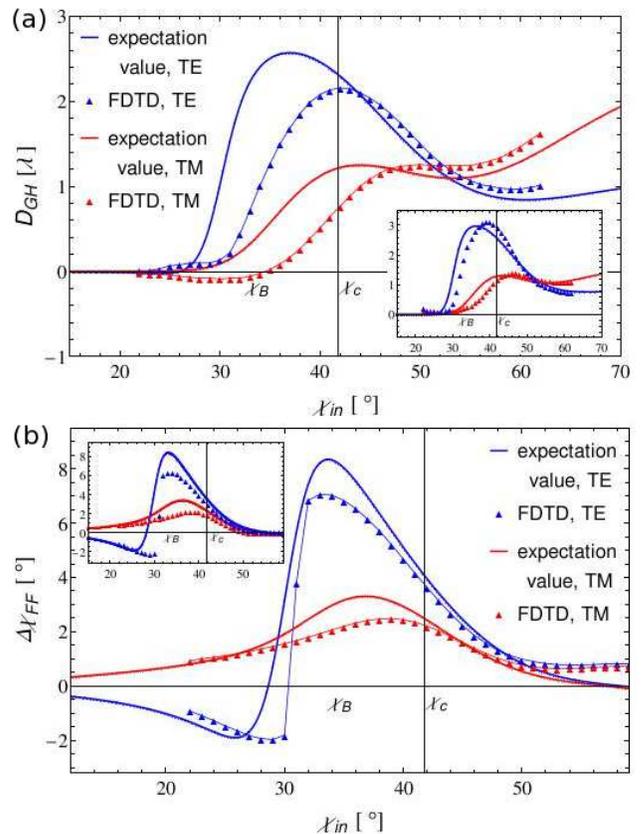} 
  \caption{\label{fig:comparison_GH_FF}(color online). Comparison between evaluation of the beam shift formulae (solid) and FDTD results (triangles) for (a) the Goos-H\"anchen shift according to Eq.~\eqref{eq:definition_GH} and (b) the Fresnel-filtering effect according to Eq.~\eqref{eq:definition_FF} of a Gaussian beam reflected at a convex interface for TM and TE polarization (red and blue, respectively); $n=1.5$, $w/\lambda=5$, $nkR=400$. 
  \emph{Insets:} Beam shifts at a planar interface. Parameters and labeling as in the main figure.}
 \end{figure}

The comparison between FDTD simulations and the expectation value approach is presented in Fig.~\ref{fig:comparison_GH_FF}, where GHS and FF are shown for both polarizations (TE in blue, TM in red) as function of $\chi_\text{in}$, the insets show the planar case. We find nice and convincing agreement between both methods without free fitting parameters.
The results coincide in the regime of total internal reflection, whereas the FDTD simulation yields smaller values than the analytical approach for subcritical angles of incidence due to two effects. First, incident and reflected beams overlap in practice, and their interference is not included in the analytical formulae. Second, the boundary discretization in the FDTD simulation induces wave scattering since the interface acts like a reflection grating. These scattered waves affect the beam propagation direction resulting in the underestimation of the beam shifts.

The GHS amounts to a few vacuum wavelengths $\lambda$, the FF effect to several degrees -- a sizable amount accompanied by a corresponding correction in the refracted light, where it will directly affect the farfield emission direction such that the total angular momentum is balanced \cite{nonHamiltonian}. 
The maximum GHS values are found around $\chi_c$, and for FF at slightly smaller $\chi_\text{in}$, in agreement with our expectations. Notice that the broad maxima originate not only in the angular composition of the incident beam, but additionally in the ``smoothed-out'' form of the curved Fresnel coefficients in comparison to their planar counterparts \cite{SchomerusHentschel_Fresnel_laws_curved}. We find for TE polarization that both the GHS and the FF are larger, by as much as a factor of two, compared to TM polarization. The polarization dependence of the reflection coefficients causes these differences. A similar behavior is known from the planar case \cite{Wolter}.
 
\begin{figure}
  \includegraphics[width=.47\textwidth]{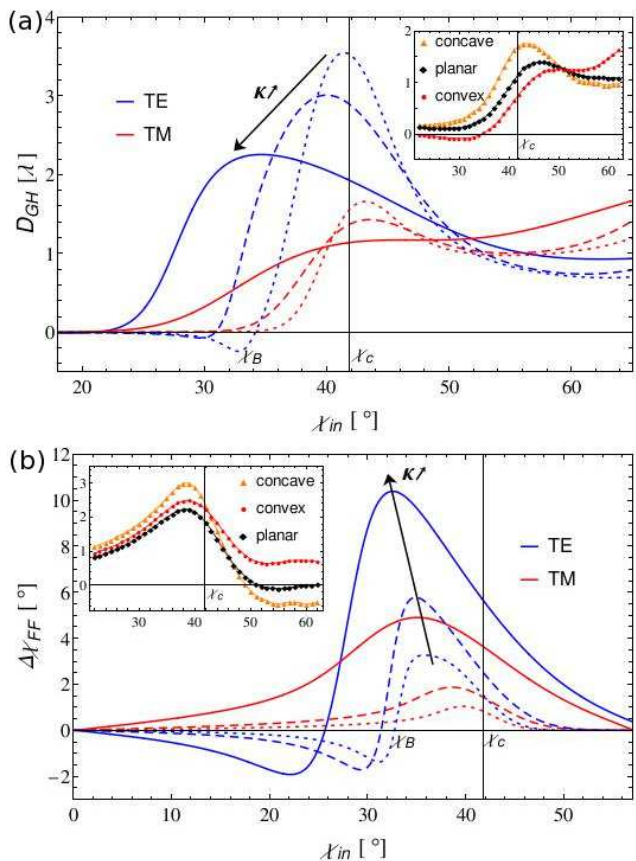} 
  \caption{\label{fig:effect_of_curvature}(color online). Effect of curvature on the beam shifts at convex interfaces. The wavenumber $k$ and the angular spread $\sigma=36$ are fixed while the curvature $\kappa=1/R$ is varied. (a) Goos-H\"anchen shift according to Eq.~\eqref{eq:definition_GH}, (b) the Fresnel-filtering effect according to Eq.~\eqref{eq:definition_FF}. TM and TE polarization in red and blue, $nkR=300$, $600$, and $900$ as solid, dashed and dotted lines, respectively.
  \emph{Insets:} Comparison of FDTD results for the beam shifts of a TM polarized beam at convex ($nkR=400$, red dots), concave ($nkR=-400$, orange triangles) and planar ($n=1.5$, black diamonds) interfaces.}
 \end{figure}

An important point concerns the curvature dependence of GHS and FF, see Fig.~\ref{fig:effect_of_curvature}. We fixed the angular spread $\sigma$ and wavenumber $k$ of the incident Gaussian beams, Eq.~\eqref{eq:gaussian}, and varied the radius $R$ of curvature only. 
We find that GHS and FF depend in opposite ways on the curvature $\kappa$: Whereas high $\kappa$ (small $R$, solid lines in Fig.~\ref{fig:effect_of_curvature}) reduces the GHS effect, FF grows considerably at the same time. The curvature dependence is significant, e.g., for TM,  
$\Delta \chi_\text{FF}$ increases with $\kappa$ from nearly zero to values comparable to those reached for TE polarization (at slightly lower $\kappa$). All maxima are further broadenend with increasing curvature, implying an importance of ray optics corrections already for $\chi_\text{in}$ well below $\chi_c$.

To understand the curvature dependence of GHS and FF, it is instructive to extend our numerical study from convex to concave interfaces, keeping the planar case as reference in between. The results are shown in the insets of Fig.~\ref{fig:effect_of_curvature}. We find the GHS to be largest at concave interfaces. 
This interesting result is easiest understood in the intuitive picture based on the effective angle of incidence $\chi_\text{in}^\text{eff}$. Assuming the penetration depth $\gamma$ to be solely determined by material properties, $\chi_\text{in}^\text{eff}$ will be smaller than $\chi_\text{in}$ for convex and larger for concave curvature. This trend is enhanced with increasing $|\kappa|$. 
Translating $\gamma$ and $\chi_\text{in}^\text{eff}$ into a GHS yields (for not too large shifts) $D_\text{GH} \approx 2 \gamma \tan \chi_\text{in}^\text{eff}$.
This explains right away our observation that GHS is largest for concave interfaces, as well as the decrease of GHS with increasing curvature at convex interfaces. 

In contrast to GHS, we find FF to be increased by any kind of curvature because the distribution of $\chi_\text{in}$ is effectively enhanced at the curved interface. As FF mainly depends on the broadness of the distribution of incident angles the results shown in Fig.~\ref{fig:effect_of_curvature}(b) are readily understood: $\Delta\chi_\text{FF}$ increases with $|\kappa|$ in the convex case (and in the concave, case not shown here) and is smallest in the planar case. 

The simplicity of the picture developed here -- compared to the high numerical cost of both full simulations and numerical evaluation of the expectation value approach -- provides an easy-to-use tool that helps to anticipate deviations from the ray picture at curved interfaces, especially for the prediction of farfield emission of microlasers. The impact of curvature becomes evident and can be either employed, e.g., for new technologies, or counteracted by, e.g., downstream optical equipment in the course of light processing. Not least we have generalized the fundamental principle of ray-wave-correspondence by a detailed discussion of how curvature affects wave-inspired semiclassical corrections. Although we dealt with the case of electromagnetic waves here, generalization to other kinds of waves is straightforward \cite{LeeKetzmerick_quantumGHS,GHSelectronic,GHSultrasonic}.

An increasing number of experiments \cite{Quasiscars,Shinohara2011,WiersigCao_ultrasmall,Tran1995,UnterhinninghofenPRE} emphasized deviations from the naive ray model and the importance of wave corrections in very small cavities, i.e., for high $\kappa$. 
Tiny fabrication imperfections can then result in undesired variations of the cavity curvature. The strong curvature dependence of FF can thus have a strong, uncontrolled effect important to know about when testing and optimizing new devices. The loss of ray path reversibility due to FF can induce a chirality (a loss of symmetry between clockwise and counter-clockwise propagating modes) in asymmetric microdisk cavitites \cite{nonHamiltonian,WiersigCao_ultrasmall}.  

Our results indicate that corrections to the ray picture in the form of the beam shift effects GHS and FF are important even when the cavity size is not too small, especially when they are affected by curvature effects. Whereas the role of GHS decreases with increasing curvature, FF becomes significantly more important at curved boundaries for both polarizations and should be taken into account when predicting the behavior and farfield properties of microoptics devices in the course of their ongoing miniaturization.

\begin{acknowledgments}
The authors gratefully acknowledge funding by the German Research Society (DFG) via the Emmy Noether Programme and the Research Unit FOR 760. We thank Daniel Kotik, J\"org G\"otte, Ulf Peschel, and Henning Schomerus for discussions and valuable advice.
\end{acknowledgments}


%

\end{document}